\documentstyle[preprint,aps,epsf]{revtex}
\begin{document}
\title{Determination of the
friction coefficient of a  Brownian particle
\\  by molecular-dynamics simulation}
\author{F. Ould Kaddour$^{1,2}$ and D. Levesque$^{1}$}
\address{$^1$Laboratoire de Physique Th\'eorique, Universit\'e Paris Sud,
B\^atiment 210, 91405 Orsay, France \\
$^2$Institut de Physique,
Universit\'{e} de Tlemcen,  BP119 Tlemcen 13000 Alg\'{e}rie} 
\date{\today}
\maketitle
\begin{abstract} 
By using the Kirkwood formula, the friction coefficient
of a solvated Brownian particle is determined from the integration 
on time of the autocorrelation function of the force that the solvent 
exerts on this particle. Extensive molecular dynamics simulations 
show that above a definite size of the studied systems
the value of the integral defining the friction coefficient
goes to a quasi constant value (a plateau) when the upper bound on
time increases. 
The minimal value of the system size 
where the integral exhibits this asymptotic
behavior, rises with  the Brownian particle size.
From the plateau, a reliable estimate 
of the friction coefficient is obtained.\\ 

\end{abstract}

In solutions, it is supposed
that the large particles  such as micelles or colloids
which  coexist with  the atoms, ions or small molecules of the
solvent behave as Brownian particles. At low concentrations
of the large particles, by using multi-scale 
analysis \cite{frie,cuk,boc1,boc2},
this hypothesis has been justified in the limit where
the ratio between the mass of the solvated  particles 
 and that of the solvent  molecules goes to $\infty$. 
It has then been established that 
the diffusion coefficient of  Brownian particles  
can be computed in term of the friction coefficient 
characterizing the force exerted on them by the solvent. When the  
Brownian particles have a quasi-macroscopic size, from hydrodynamic
arguments  the Stokes \cite{lan} law can be derived. It gives an expression of 
the friction coefficient $\xi=C R\eta$ where $R$ is the size of 
the particles, $\eta$ the viscosity of the solvent and $C$ a numerical 
coefficient depending on the possible choices of boundary conditions 
at the interface between solvated particles and solvent.

However in many suspensions, such as
ionic solutions, the values of ratios of masses 
and sizes between the solvated  particles and solvent molecules 
are only of the order of 10 and
the possibility that the solvated particles can be considered
as Brownian particles becomes questionable.
Several theoretical works \cite{boc3,sari} and works based on numerical
simulations \cite{lag,bre,esp,oul} have been devoted to this question. 
The main problem addressed in these works was that of
the determination of the lower bounds of the size and mass  
ratios above which, to a good approximation,
the motion of the solvated particles is Brownian.  
The criterion chosen to locate these bounds
was that the diffusion coefficient of the solvated particles obeys
to the relation between the diffusion coefficient $D$
and friction coefficient $\xi$ 
strictly valid only for brownian particles $
D=k_B T/\xi $
($k_{B}$ is Boltzmann's constant, $T$ the temperature
of the solvent).
The main concern, when  the Stokes estimate  
of $\xi$ is used, is the choice of the hydrodynamic
 boundary condition between solvated particles and solvent which is
 well defined only when the solvated particle has a macroscopic size. 
This last shortcoming can be overcome by computing $\xi$
from its exact expression for a Brownian particle
derived by Kirkwood \cite{woo} and later, more rigourously,
from multi-scale analysis \cite{cuk,res}.
Obviously this method seems the correct way to proceed in order
to check the brownian behavior of a solvated particle. However,
as it was discussed in the literature \cite{lag,esp,woo,sud,hel,mo}, 
this method is not easy to use in simulations
 due to important finite size effects. 
This work is devoted to discuss this problem and to establish 
in what conditions, in a numerical simulation, the friction 
coefficient of a solvated particle of large mass and  size can be 
credibly determined.

The friction coefficient $\xi$ is given in terms of the integration
on time $t$ of the equilibrium autocorrelation function $<{\bf F}(0) \, . \, {\bf F}(t) >$
of the instantaneous microscopic force ${\bf F}(t)$ experienced by 
the Brownian particle :
\begin{equation} \xi= \frac{1}{3k_{B} T} \int_{0}^{\infty}
<{\bf F}(0) \, . \, {\bf F}(t)> \, dt \, .\end{equation}
This expression of the friction coefficient has
the same form as the Green-Kubo relations
used to calculate the transport coefficients.

For finite size systems, the computation  of $\xi$ from Eq. (1) 
and, more generally, that of the transport coefficients from
Green-Kubo relations are confronted
with a problem that we illustrate for this specific case. From
the momentum conservation, the force ${\bf F}(t) $ acting on one Brownian particle
in a solvent of $N$ molecules is  given by 
\begin{equation} {\bf F}(t)= - {\frac{d \, {\bf P}(t)}{dt}} \equiv -{{\bf {\dot{P}}}(t)}
= - \sum_{i=1}^{N} m \frac{d\,{\bf v}_{i}(t)}{dt} \, ,
\end{equation}
where $m$ is the mass of the solvent molecules with velocities ${\bf v}_{i}(t)$
($i=1, ...,N$) and
$\xi_N$ can be written as
\begin{eqnarray} \xi_{N} & = &  \lim_{t \rightarrow \infty} 
\xi_{N}(t)  \\
 &=&   \frac{1}{3k_{B} T} \lim_{t \rightarrow \infty} \int_{0}^{t}
<{\bf F}(0) \, . \, {\bf F}(\tau ) >_{N} \, d\tau \nonumber \\
 & = & \frac{1}{3k_{B} T}  \lim_{t \rightarrow \infty, \, s \rightarrow
 \infty } \int_{0}^{t} d\tau
 \frac{d}{d\tau} {\frac{1}{s}}\int_{0}^{s} 
 {\bf F}(u).  {\bf P}(\tau+u )  \, du \nonumber \\
& = & - \lim_{t \rightarrow \infty } {\frac{ < {\bf \dot{P}} (t). {\bf P}(0) >_{N}} 
 {3k_{B} T} }+ {\frac{ < {\bf \dot{P}} (0) . 
 {\bf P }(0) >_{N} }{3k_{B} T} }.  \end{eqnarray}

The last term of Eq. (4) vanishes by symmetry  on time and 
it is expected that, for $t  \rightarrow \infty $,
the first term also vanishes due to the loss of correlations
between ${\bf \dot{P}} (\infty)$ and $\, {\bf P}(0)$, according to the
ergodic postulate of the equilibrium  statistical mechanics,
with the final result  that $\xi_N$ should be zero. 
The well known way to overcome this paradox
is that the thermodynamic limit on $N$ must be taken before that
the limit $t  \rightarrow \infty $ is performed. The argument can
be summarized by guessing  that at large values of $t$ and $N$,
$\xi_{N}(t)$ can be written in the form $c \, g(at/N)$ where $c$ and $a$ are 
coefficients independent of $N$ 
and $g(at/N)$ is a decreasing function of $t$ equal to $0$ at $t=\infty$ 
and normalized to $1$ at $t=0$.
By perfoming the thermodynamic limit $N \rightarrow \infty$ before
the limit $t \rightarrow \infty$,  $\xi$ is given by $\lim_{t \rightarrow \infty}
\{ \lim_{N \rightarrow \infty } \xi_{N}(t) \}= c $ and $\xi$ is now
finite. 
 
 If in simulations $N$ is large enough,
it can be expected, following the remark
made by Kirkwood in \cite{woo}, that, in the range
of values of $t$ where $\xi_{N}(t)$ reaches its asymptotic
form $\simeq c \, g(at/N) $, $t$ is such as $t<< a/N$. In this domain
of $t$, $\xi_{N}(t)$ is given by $\xi_{N}(t) \simeq c + acg'(0)t/N$ and 
presents a quasi plateau or a  slow linear decay with $t$ from 
which the value of $\xi$ in the thermodynamic limit can be estimated.

As it was proposed, for instance, in \cite{boc3,esp} it is possible to give
a specific analytic form to $g(at/N)$ by supposing that,
following  the Onsager's principle, the regression of the fluctuations 
of ${\bf F}(t)$ at large $t$ is governed by the laws of the
macroscopic hydrodynamics. According to these laws, the force exerted by
the solvent on the Brownian particle is proportional to the momentum 
${\bf P}(t)$ carried by the solvent, i.e.

\begin{equation} {\bf F}(t)=  {\frac{\xi_o}{Nm}}{\bf
P}(t) \, \end{equation} 
a relation which implies that
\begin{equation} <{\bf P}(t) \, . \, {\bf P}(0)>_{N}= 3Nmk_{B}T \exp(-\frac{\xi_o}{Nm}
t)\end{equation}
and then
\begin{equation}
 \xi_{N}(t) = {-\frac{<{\bf \dot{P}}(t) \, . \, {\bf P}(0)>_{N}}{3k_{B} T}}=\xi_o
\exp(-\frac{\xi_o}{Nm}t).\end{equation}

At large $N$ and $t$ such as $t<< \xi_o /Nm$, an expansion of the exponential yields
to a linear expression, similar to that of $\xi_{N}(t)$ given above, 
allowing to determine the  friction coefficient as
\begin{equation}  \xi_{N}(t) 
 =\xi_o \, (1- \frac{\xi_o}{Nm}t+ \, ... \, ) \,.\end{equation}

  In order to investigate the possibility of a computation
of $\xi$ following the procedure describes above,
 we have realized a set of molecular-dynamics 
simulations with increasing values of $N$.

The studied systems are made of $N$ molecules of solvent enclosed 
in a periodic cubic box of volume $V$. In this box, 
a particle is immersed and supposed to have
a size and mass $M$ large compared of those of the solvent molecules.
Hence the mass of this particle satisfies to the condition 
required so that the relation between $D$ and $\xi$, given above, applies. 
When $M$ is  large, it is possible to consider that, 
in the time scale accessible in a simulation, the particle is immobile.

The molecules and the fixed particle interact
through a Lennard-Jones (LJ) potential modified with a cubic spline, 
as describe  in a previous paper
\cite{oul}. This potential has  the form $ v_ {ij}(r) = \epsilon_{ij}
 f({r}/{\sigma_{ij}})$ where the indices $i,j=1$ or $2$ refer to the
solvent and  fixed particle, respectively. The parameters $\sigma_{ij}$
are such as $\sigma_{12}=(\sigma_{11} + \sigma_{22})/2$, and $\epsilon_{ij}$
are equal $\epsilon_{12}=\epsilon_{22}=\epsilon_{11}$.  The unit of time  is 
chosen equal to $\tau_0 ={\sqrt{(m\sigma_{11}^2/\epsilon_{11})}}$ and that of 
energy, length, and mass are chosen, respectively, equal to $\epsilon_{11}$,
$\sigma_{11}$, and $m$.  The values of the solvent density and temperature are
 $\rho^*\equiv N \sigma_{11}^3/V \simeq0.84$ and $T^* \equiv k_BT/\epsilon_{11}=1.0$, 
 thus specifying a dense liquid state near the triple point of the LJ system.
The simulations were realized at constant energy using the standard 
Verlet algorithm  \cite{all}, with a time step  $\Delta t=0.005 \, \tau_0$. 
Typical simulation runs are  carried out for 20000 equilibration time steps 
followed by 4 to 8 millions time steps. During the runs,
the time autocorrelation function of ${\bf F}(t)$ is  computed over a sequence 
of blocks of $4000\Delta t$  to allow an evaluation of statistical errors. 
We have considered two different sizes for the Brownian particle, namely
$\sigma_{22}=4.0$ and $7.0$ and systems of increasing values of $N$ : $864, \, 1500,\, 
5324, \, 12000, \, 32000$ and $55296$. In order  to maintain constant the value of the 
pressure, the volume of the simulation box was slightly increased, 
when $ \sigma_{22}$ was varied from $4.0$ to $7.0$.

 We first discuss the case of $\sigma_{22}=4.0$. In Fig. 1, we show $\xi_N(t)$ 
 as a function of reduced time $t/\tau_0$. When
 $N$ is increased from $864$ to $32000$, the behavior of $\xi_N(t)$
 changes drastically in the domain of $t/\tau_0 > 5.0$. For the low values
 of $N$, $\xi_N(t)$ goes rapidly to zero and for the two larger values
 of $N$, it is almost constant. Qualitatively, this behavior of $\xi_N(t)$
 at large time corresponds to that expected when $N$ increases. In particular,
 if this behavior is described by  Eq. (7), the results
 presented in Fig. 1 can be interpreted as the transition
 between the exponential decrease  of $\xi_N(t)$ given by Eq. (7) and 
 the linear decrease, at large $N$, given by Eq. (8).
 Quantitatively, it should 
 be possible to obtain
 the value of $\xi$ from the fit of $\xi_N(t)$, for $t/\tau_0$ between
 $12$ and $20$, to an exponential form when $N$ is $\simeq 1000$ and
 a linear form when $N$ is $\simeq 20000$.

  However, the possibility
 that the values of $\xi$, determined from the fits made at  different
 values of $N$, coincide within the statistical uncertainties,
 supposes that finite size effects, in particular those
 associated to the use of the periodic boundary conditions, 
 do not affect the long time behavior of $\xi_N(t)$.
 From the fit  of the multiplicative constant
 of the exponential (cf. Eq. (7)) the estimates of $\xi$ are 
 : $219.4$ ($N=864$), $144.7$ ($N=1500$) and $122.9$ ($N=5324$)
 and from that of the constant term in the linear form
 (cf. Eq. 8) they are :  $113.3$ 
($N=12000$) and $110.7$ ($N=32000$). Clearly, for the two large
systems these values of $\xi$ agree within the statistical error
equal to 10-15\%. But the factor of 2 between
the values found at $N=864$ and $N=32000$
 must be attributed to finite size
effects. The side lenghts $L$ of the simulation 
cells being for these two values
of $N$ equal to $\sim 10$ and $\sim 35$, due to the periodic
boundary conditions the fixed particle is distant
from these nearest replica by the same lengths. 
The sound velocity $c_s$ of the solvent, for the considered
thermodynamic state, being in reduced units $\sim 6$,
gives typical times $L/c_s$ of $1.5$ and $6$ beyond of which 
the finite size effects resulting from the mutual influence 
between the fixed particle and its replicas can affect 
the correlation function. The magnitude of these effects 
on the value of $\xi$ is difficult to assess. It has been 
quantitatively discussed only for the bulk values of 
the transport coefficients of diffusion or viscosity 
in dense fluids \cite{erpe}, corrections of about $\sim 10 \%$  
have been found for systems of $N\sim 1000$ molecules
and they seem much larger on $\xi$.
Other estimates of $\xi$ are obtained from
the fit of the coefficient of $t$ in
the exponential (Eq. (7)) or the linear
approximation (Eq. (8)).
For the small values $N$, we obtained $205.5$ ($N=864$)
 and $143.3$ ($N=1500$), for the large values of $N$, 
$98.4$ and $91.1$. These results seem to confirm that
the asymptotic behavior of $\xi_N(t)$ is well
described by Eq. (7) taking into account the finite 
size effects.
  
\begin{figure}
\epsfxsize=4.5 truein
\epsfbox{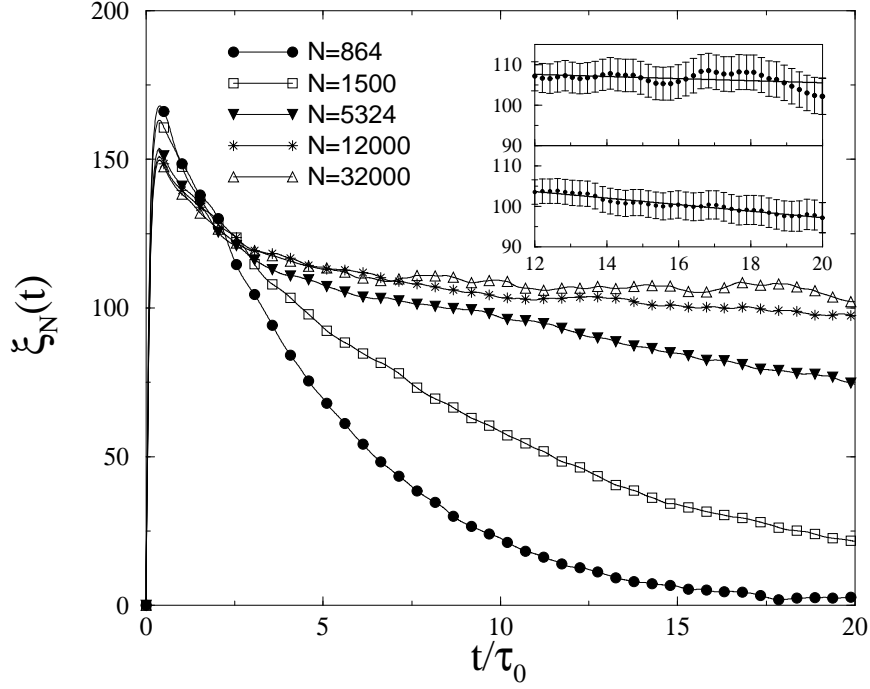}
 \caption{ $\xi_N(t)$ for the fixed particle of size $\sigma_{22}=4.0$
 and increasing values of $N$. Insert : asymptotic behavior with error bars
and its linear fit at $N=12000$ and $32000$}
\label{Fig. 1}
\end{figure}
 
 \begin{figure}
\epsfxsize=4.5 truein
\epsfbox{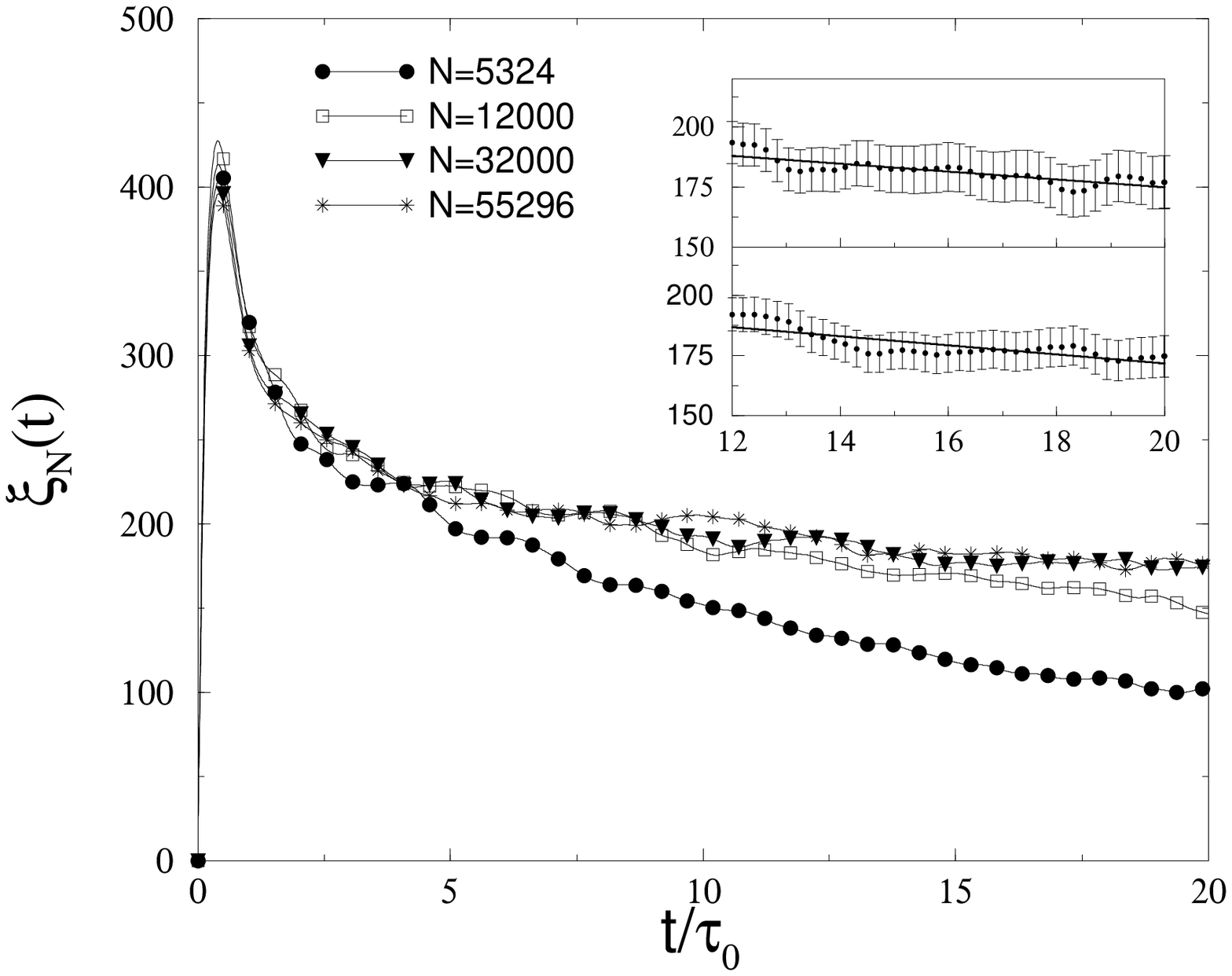}
 \caption{ $\xi_N(t)$ for the fixed particle of size $\sigma_{22}=7.0$
 and increasing values of $N$. Insert : asymptotic behavior wih error bars and
 its linear fit for $N=32000$ and $55296$}
\label{Fig. 2}
\end{figure}

We consider now the case
of the fixed particle with a size $\sigma_{22}=7.0$.
The $\xi_N(t)$ functions are plotted
in Fig. 2. for $N$ equal to $5324$, $12000$,
$32000$ and $55296$. The behavior of  $\xi_N(t)$,
when $N$ increases, is similar to that obtained 
with the particle  of size $\sigma_{22}=4.0$, but it is only for $N\ge32000$ 
that $\xi_N(t)$ presents a slow linear decrease
for $t/\tau_0 > 10.0$. For the particle of size $\sigma_{22}=4.0$, 
we remark that this latter behavior is reached
for systems smaller by a factor $\sim 2 $ and, then,
the size of the fixed particle has an important
influence on the value of the system size
where $\xi_N(t)$ exhibits a slow linear decrease at large time.
This remark is confirmed by the results of
the fits of $\xi_N(t)$ by an exponential at $N=5324$ and $12000$ 
or a linear function at $N=32000$ and $55296$. The values of $\xi$
are $217.0$,  $244.2$, $209.3$ 
and $207.5$ respectively. We notice
that, the linear behavior of $\xi_N(t)$, at large $t$, 
corresponds obvioulsy to the fact that
$<{\bf F}(0) \, . \, {\bf F}(t) >$ decreases very slowly
in the same domain of $t$. This point is illustrated in Fig. 3
where this behavior is clearly seen for $N=12000$, $32000$ and $55296$
within statistical errors.

 \begin{figure}
\epsfxsize=4.5 truein
\epsfbox{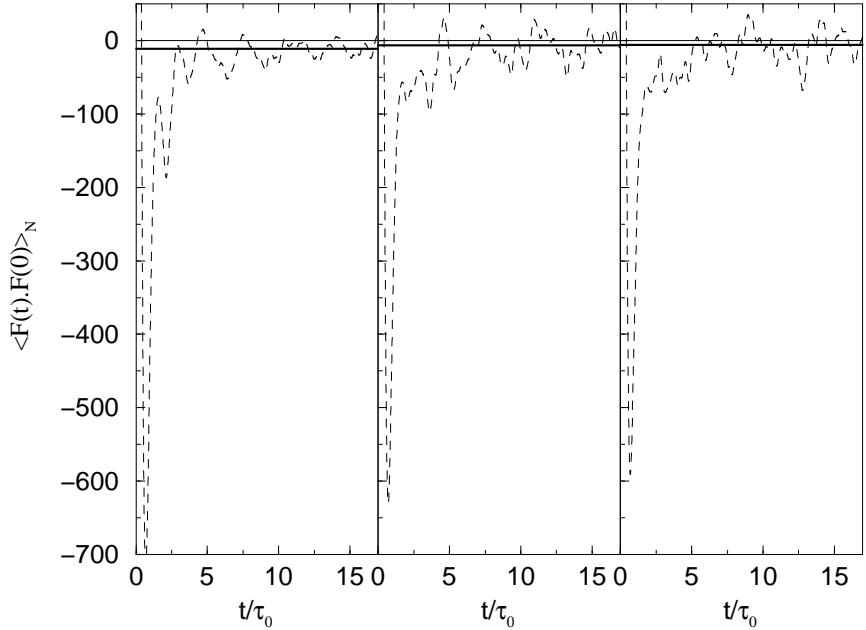}
 \caption{Autocorrelation function $<{\bf F}(0) \, . \, {\bf F}(t) >_N$
for the particle of size $\sigma_{22}=7.0$ at $N=12000$, $32000$
and $56296$. Solid lines : estimate of its almost constant 
value from its average for $t/\tau_0>12.0$ and $<20.0$ :
$\sim$ -12.6, -6.6 and -6.1.}
\label{Fig. 3}
\end{figure}

 From the results of the present simulations it seems
 needed to adopt a critical point of view on the
 previous works made in order to determine the friction
 coefficient of a brownian particle 
 and to check when a particle of large masses
 and sizes can be considered as a brownian
 particle. The most important criticism  is that,
 in these works the sizes of
 the system studied in the simulations were 
 too small. This remark applies, for instance,
 to the simulations presented in \cite{boc3}  where 
 the friction coefficient of a fixed hard sphere of
 diameter $4d$ in a solvent made of hard spheres of diameter $d$
 was computed at a density $\rho d^3 \simeq 0.471 $.
 Such a computation corresponds closely 
 to that made in this work for a fixed particle with 
 $\sigma_{22}=4.0$. In \cite{boc3},
 the largest considered system had a size of $N=1500$
 which, as discussed above, seems too small to obtain
 a good estimate of $\xi$ from an exponential fit
 of $\xi_N(t)$ at large times and, then, to check
 the validity of Stokes estimate of $\xi$. The system size 
 used in \cite{esp} for the computation of the friction coefficient 
 of a fixed particle in a LJ type system being of the order of $N= 1000$,
 finite size effects should also affect the simulation data.
 
 As mentionned already, the asymptotic form of $\xi_N(t)$
 at large times has been discussed in many works
 in the literature, for instance in \cite{sud}, \cite{hel}
 and \cite{mo}, in particular the question of 
 the occurence of a domain of time where
 the friction coefficient $\xi_N(t)$ should exhibit
  a plateau. It has been proposed in \cite{lag} to bypass 
 the search of such a plateau in  $\xi_N(t)$ by computing 
 $\xi$ from the integration of $<{\bf F}(0) \, . \, {\bf F}(t) >$
 from $t=0$ to the value of $t=t_1$ 
 where, for the first times when $t$ increases,
 $<{\bf F}(0) \, . \, {\bf F}(t_1) >$ becomes $0$. 
In our simulations $t_1/\tau_0 \simeq 0.5$,
 clearly from the comparison between Fig. 2 and Fig. 3 of
 $\xi_N(t)$  and $<{\bf F}(0) \, . \, {\bf F}(t) >_N$,
 such a method to estimate $\xi$ seems problematic
 since the value of $\xi_N(t_1)$ for instance at $N=55296$
 does not agree with the value $\xi$ obtained from the
 analysis of the asymptotic behavior of $\xi_N(t)$.
 
 Since the present simulations show that system sizes
 of  $N\simeq20000$ are needed to  correctly estimated $\xi$, 
 it is expected that similar system sizes are needed
 to compute  $D$ in order to avoid finite size effects. 
 For instance in \cite{oul} for $N=5324$, for a brownian particle with 
 $\sigma_{22}=4.0$ and $M=60$ in a LJ solvent at a thermodynamic state identical
 to that considered here, it was found  $D=0.077$. By using $D=k_B T/\xi $
 this value of $D$ agrees well with that of $\xi\simeq120$ obtained 
 in this work  at $N=5324$, but it differs by $10\%$ from 
 that, computed at $N=32000$, $\xi\simeq110$. A new determination
 of $D$ at $N=32000$ gives $D=0.095$, which now agrees with
 our estimate of $\xi$ at the same value of $N$ from the slow linear 
 decay of $\xi_N(t)$ for $t/\tau_0>12$.
 
  In conclusion from our simulations realized for increasing system sizes
  from $N\sim 1000$ to $\sim 56000$, we have shown that 
  the qualitative behavior of $<{\bf F}(0) \, . \, {\bf F}(t) >$ and, 
  consequently, that of $\xi_N(t)$ is in excellent agreement 
  with that guessed by Kirwood \cite{woo} and in subsequent works.  
  \cite{hel,mo}. For large times, the representation of $\xi_N(t)$ by
  Eq. (7), derived from a simple argument based on
  Onsager principle, seems an adequate model of this asymptotic
  behavior. From a quantitative point of view, in spite
  of simulation runs totalizing 4 to 8 millions of
  time steps the statistical uncertainties on
  the friction coefficient stay of the odrer of $15$\%,
  a reduction of this uncertainty by a order of magnitude
  seems  beyond what it is possible to make by using present
  standard computers. We stress that the aim of this work
  was the investigation of the variation of $\xi_N(t)$
  with $N$. In the thermodynamic limit, the asymptotic behavior
 of $\xi_N(t)$ should be an algebraic decay at  very large time.
 It was not considered here, in particular, because its
 amplitude is much smaller than the present uncertainties
 on $\xi_N(t)$ in this domain of time \cite{oul,erpe}.

\end{document}